\documentclass[twocolumn,aps,prl,superscriptaddress,showpacs]{revtex4-1}
\usepackage[utf8]{inputenc}
\usepackage{color}
\usepackage{amsmath}
\usepackage{amssymb}
\usepackage{graphicx}
\usepackage{esint}
\usepackage[unicode=true,
 bookmarks=false,
 breaklinks=true,pdfborder={0 0 0},backref=false,colorlinks=true]
 {hyperref}

\makeatletter

\usepackage{epstopdf}\usepackage{subfigure}
\usepackage{float}
\usepackage{amsfonts}
\usepackage{bm}
\usepackage{subfigure}
\usepackage{braket}

\makeatother
\usepackage{tikz}

\begin{document}

\title{Unique surface-state connection between Weyl and nodal ring fermions in ferromagnetic material Cs$_{2}$MoCl$_{6}$}

\author{Tiantian Zhang}
\thanks{These two authors contributed equally}
\email{zhang.t.ac@m.titech.ac.jp}
\affiliation{Department of Physics, Tokyo Institute of Technology, Ookayama, Meguro-ku, Tokyo 152-8551, Japan}
\affiliation{Tokodai Institute for Element Strategy, Tokyo Institute of Technology, Nagatsuta, Midori-ku, Yokohama, Kanagawa 226-8503, Japan}

\author{Daisuke Hara}
\thanks{These two authors contributed equally}
\affiliation{Department of Physics, Tokyo Institute of Technology, Ookayama, Meguro-ku, Tokyo 152-8551, Japan}

\author{Shuichi Murakami}
\affiliation{Department of Physics, Tokyo Institute of Technology, Ookayama, Meguro-ku, Tokyo 152-8551, Japan}
\affiliation{Tokodai Institute for Element Strategy, Tokyo Institute of Technology, Nagatsuta, Midori-ku, Yokohama, Kanagawa 226-8503, Japan}

\begin{abstract}


For topological materials with coexistence of Weyl nodes and nodal rings, the surface-state configuration and connection are unique yet have never been studied and discussed before. 
In this paper, we predict a ferromagnetic (FM) material, Cs$_{2}$MoCl$_{6}$, with coexistence of Weyl and node-ring fermions in its spinful FM electronic band structure, which is unusual since FM materials are very rare in nature and node-ring band crossings will usually open a gap when spin-orbit coupling (SOC) is taken into consideration. 
We find that the surface states of Cs$_{2}$MoCl$_{6}$ show different properties along different directions, i.e, the surface states are in the drumhead shape showing the node-ring property on the (001) surface and in the helicoid shape showing the Weyl property on the (010) surface. 
Interestingly, both the drumhead surface states and the helicoid surface states will cross the projected points of the Weyl and nodal ring along different directions. 
In particular, helicoid surface states on the (010) surface will meet the nodal ring tangentially, with their shapes change abruptly as a function of the energy. 
We implement both first-principle calculation and an analytical model to understand the unique surface-state connection for systems with the coexistence of Weyl nodes and nodal rings (or nodal lines). This result is universal and irrespective of the presence/absence of and time-reversal symmetry ($\mathcal{T}$).

\end{abstract}

\maketitle

\paragraph*{Introduction}

In the recent years, topological magnetic materials \cite{zou2019study,nie2017topological,PhysRevLett.124.016402,PhysRevLett.117.236401,PhysRevLett.124.076403,HgCr2S4e,prlqahe,belopolski2019discovery,chang2013experimental,kuroda2017evidence,kim2018large,wang2018large,belopolski2019discovery,li2020giant,PhysRevLett.122.057205,morali2019fermi,liu2019magnetic,xu2020high} draw much attention due to their exotic properties, $e.g.$, topological axion insulator states \cite{PhysRevLett.122.206401,xu2019higher,li2019intrinsic}, intrinsic quantum anomalous Hall effect without external magnetic fields \cite{HgCr2S4e,prlqahe,chang2013experimental,kim2018large,li2020giant,wang2018large}, and the display of magnetism, topology, spin dynamics and quantum transport. 
However, experimentally synthesized and verified magnetic topological materials are very limited, especially for the magnetic topological semimetals \cite{belopolski2019discovery,kuroda2017evidence,kim2018large,wang2018large,belopolski2019discovery,li2020giant,PhysRevLett.122.057205,morali2019fermi,liu2019magnetic}. 
Furthermore, compared to the bulk topology, properties of topological surface states like the configuration and connection are not so well understood in some systems. 
In this paper, we propose a ``high-quality'' FM material Cs$_{2}$MoCl$_{6}$ \cite{hu2005crystal} with space group $Fm\bar{3}m$ that have both Weyl and node-ring fermions in its spinful electronic band structure. 
We find unique surface-state connection between the Weyl and node-ring fermions along different surface directions by both DFT and model analysis due to an intriguing interplay between drumhead surface states (DSSs) from nodal rings and helicoid surface states (HSSs), i.e., Fermi-arc surface states, from Weyl nodes. 
We show that the DSSs always go across the Weyl nodes, when the projection of the Weyl nodes are inside that of the nodal ring. We also show that how the HSSs connect the projection of the nodal ring and the curvature of HSSs depend sensitively on the energy. 
{Such unique surface-state connection also appear in several previous studies, yet lack of discussion} \cite{song2020symmetry,zhao2021coexistence,xu2011chern,sun2018conversion,peng2020topological}. Our results show that the connection pattern of the surface states is surface-dependent for systems with both Weyl and node-ring fermions, and it is universal for systems with/without considering SOC and $\mathcal{T}$ symmetry.


\begin{center}
\begin{figure}
\includegraphics[scale=0.86]{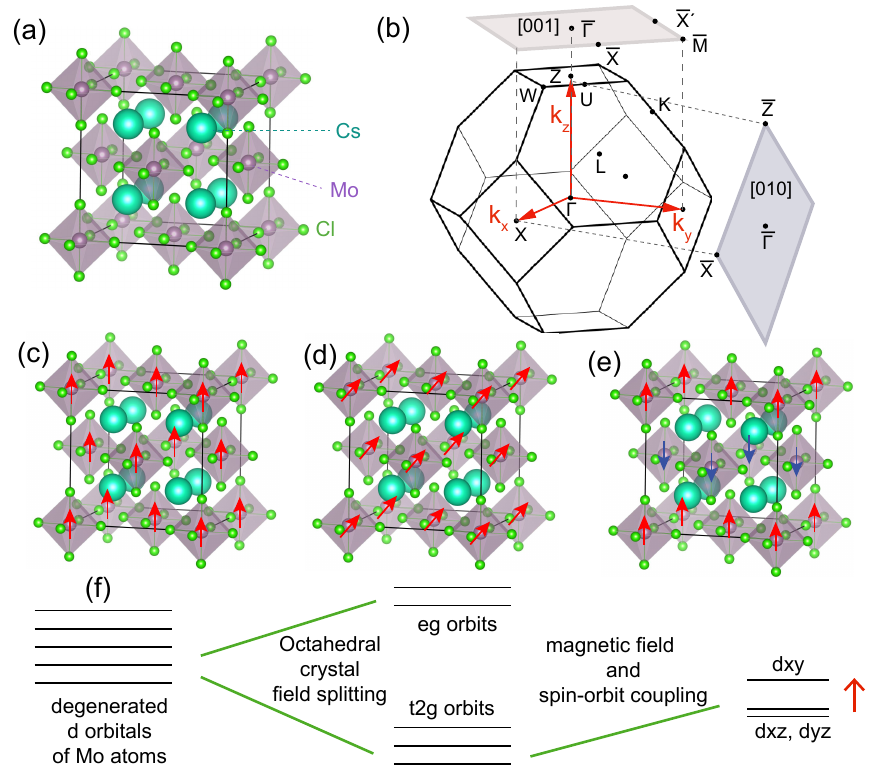}\caption{Crystal structure, Brillouin zone and splitting of $d$ orbitals of Cs$_{2}$MoCl$_{6}$. (a-b) are the crystal structure and (surface) Brillouin zone of Cs$_{2}$MoCl$_{6}$. (c-e) are three different types of magnetic structure of Cs$_{2}$MoCl$_{6}$ with ferromagnetic phase along [001] direction, [111] direction and antiferromagnetic phase, respectively. (f) shows the splitting of $d$ orbitals under different fields. 
\label{fig:FIG1}}
\end{figure}
\end{center}

\paragraph*{Crystal and magnetic structure for Cs$_{2}$MoCl$_{6}$}

Cs$_{2}$MoCl$_{6}$ belongs to space group \#225, and the lattice constant we use in this paper is $a$ = 10.57 $\AA$ after a relaxation on atomic positions, which is slightly different from the experimental data 10.21 $\AA$ at room temperature \cite{hu2005crystal}. For most of the compounds having a similar crystal structure with Cs$_{2}$MoCl$_{6}$ ($e.g.$, replacing Cs, Mo and Cl by other alkali metal elements, transition metal elements and halogen elements, respectively), their magnetic properties have not been explored by experiments except for Cs$_{2}$CoF$_{6}$ and Rb$_{2}$CoF$_{6}$, which have a FM phase at low temperature \cite{quail1972complex}, and Cs$_{2}$MoCl$_{6}$ is also among the ones lacking of exploration. 

In order to explore the magnetic structure of the ground state in Cs$_{2}$MoCl$_{6}$, we calculate the total energy of the systems with three different magnetic structures for \#225 by DFT, which are shown in Fig. \ref{fig:FIG1} (c-e). 
The total energy for the FM phase in Fig. \ref{fig:FIG1} (c) and (d) is about 20 meV lower than the antiferromagnetic phase in Fig. \ref{fig:FIG1} (e) per unit cell, and the FM phase with magnetic order on the (001) surface (Fig. \ref{fig:FIG1} (c)) will have a smaller total energy about 1 meV per unit cell than that along [111] direction (Fig. \ref{fig:FIG1} (d)). Thus, we will choose the one with magnetic moment along the highest symmetry direction, i.e., [001] direction to explore the topological properties in Cs$_{2}$MoCl$_{6}$. 
In the following, we will introduce topology of the electronic band structure of Cs$_{2}$MoCl$_{6}$ both in the nonmagnetic (NM) phase and in the FM phase separately.

\begin{center}
\begin{figure}
\includegraphics[scale=0.6]{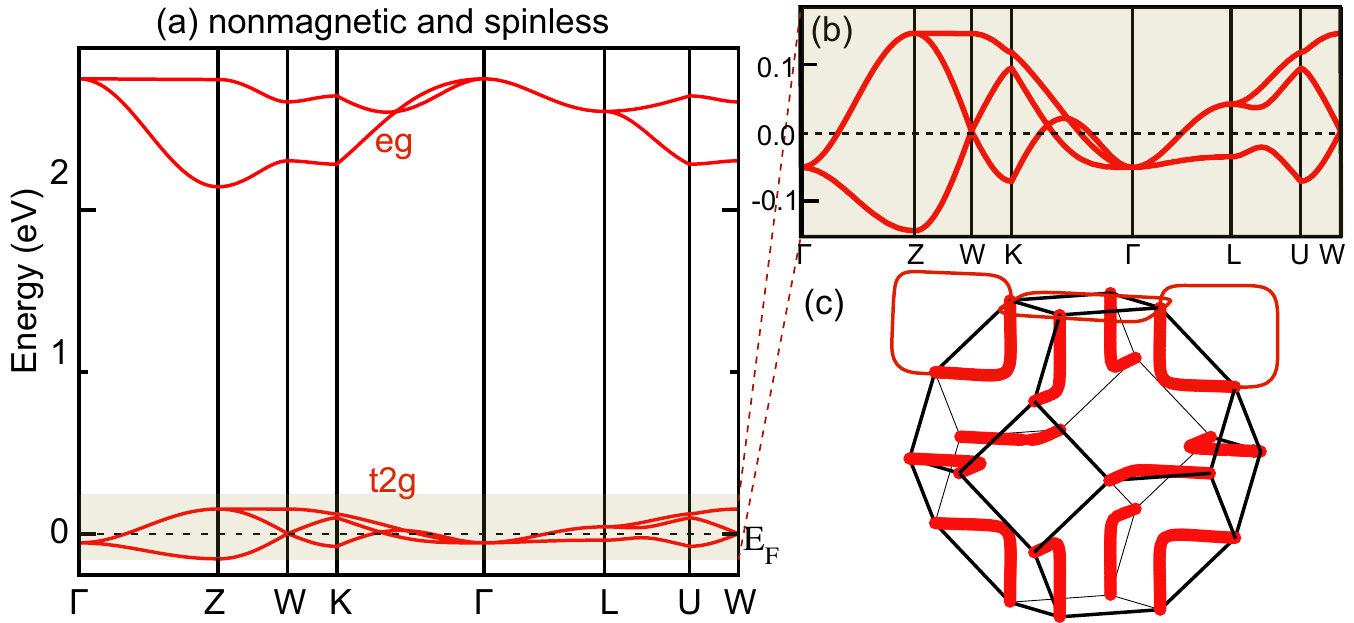}\caption{Spinless electronic band structure and the nodal chain distribution of Cs$_{2}$MoCl$_{6}$ in the NM phase. (a) is the spinless electronic band structure of Cs$_{2}$MoCl$_{6}$ in the NM phase, where the $t_{2g}$ orbits are magnified in (b). (c) shows the nodal chain band crossing formed by the lowest two bands of $t_{2g}$ orbits. 
\label{fig:FIG2}}
\end{figure}
\end{center}

\paragraph*{Band structure of Cs$_{2}$MoCl$_{6}$}
In the spinless NM electronic band structure, $d$ orbitals from $Mo$ will split into $e_{g}$ and $t_{2g}$ orbitals due to the octahedral crystal field environment and the splitting is very huge due to the large $p$-$d$ hybridization of $Mo$ and $Cl$, as shown in Fig. \ref{fig:FIG1} (f) and Fig. \ref{fig:FIG2} (a). Thus, we only need to study the $t_{2g}$ orbitals located near the Fermi energy and they exhibit nontrivial topologies, i.e., nodal chains formed by the two lowest $t_{2g}$ orbitals crossing the Fermi level. 
Due to the existence of mirror symmetries, all the nodal rings will be pinned on the $M_{x}$, $M_{y}$ and $M_{z}$ planes, and meet with each other at the high-symmetry point $W$, forming a nodal chain semimetal, as shown in Figs. \ref{fig:FIG2} (b-c).

\begin{center}
\begin{figure}
\includegraphics[scale=0.8]{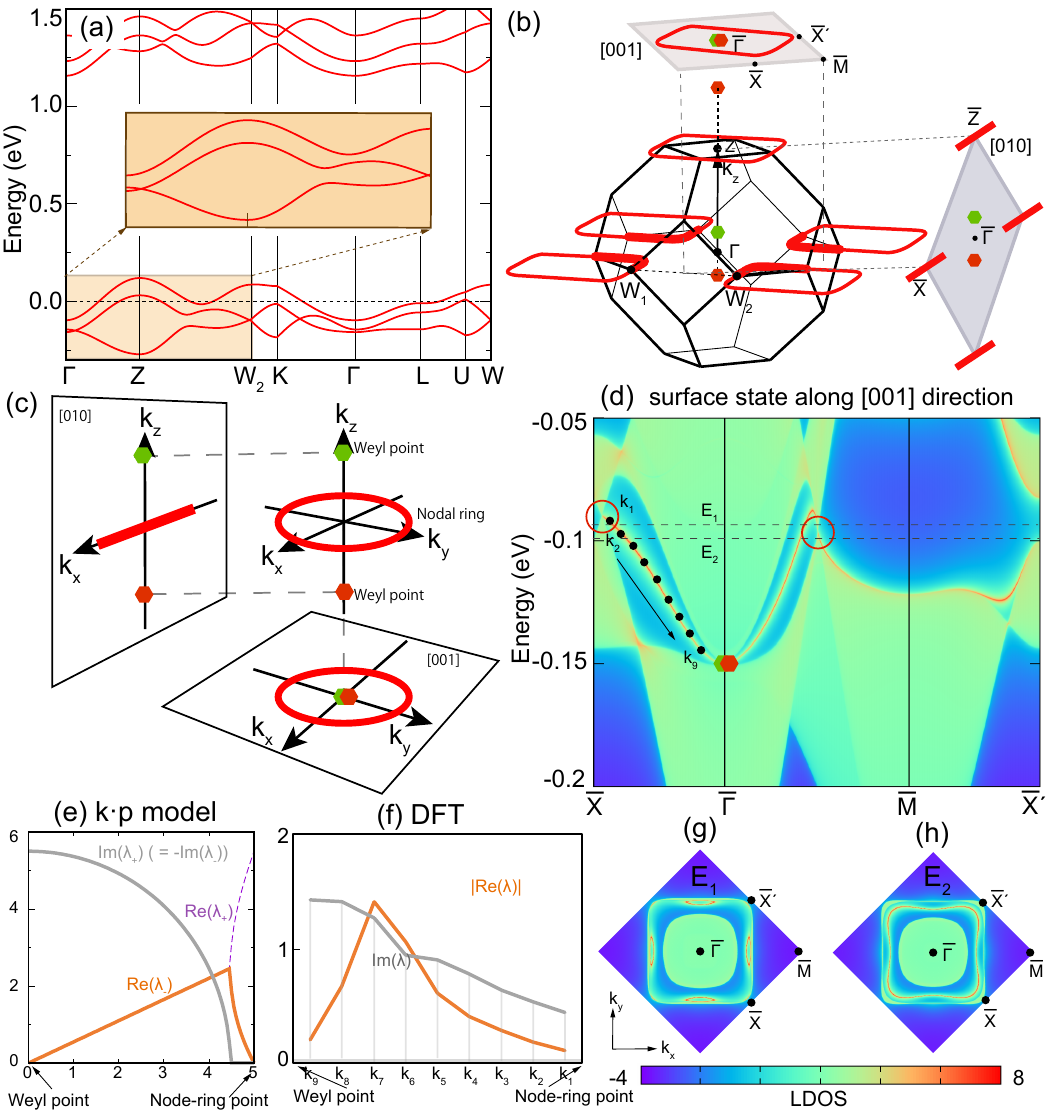}\caption{Spinful electronic band structure and topological calculations on Cs$_{2}$MoCl$_{6}$ in the FM phase. (a) is the spinful electronic band structure of Cs$_{2}$MoCl$_{6}$ in the FM phase with magnetic moment along [001] direction. (b) show the positions of a pair of Weyl points (red and green hexagons) and nodal rings in the bulk, (001) and (010) surface BZ.
(c) Schematic figure of the projected positions of the nodal ring and Weyl points on the (001) and (010) surfaces. 
(d) is the surface states, together with the projection of the Weyl points (red and green hexagons) and the nodal ring (red circles) on the (001) surface.
{(e) is the real and imaginary parts for $\lambda_{+}$ obtained by the $k\cdot p$ model with parameters of $\tilde{D}$ = 1 and $\sqrt{m}$ = 5, where Im($\lambda_{-}$) = $-$Im($\lambda_{+}$). }
(f) is the absolute value of the real part $|$Re($\lambda$)$|$ and the imaginary part Im($\lambda$) obtained by the DFT calculation, where the $k$ points are labelled in (d).  
(g-h) are the DSSs with different energies labelled in (d), where $E_{1}$ corresponds to the energy of nodal rings.
\label{fig:FIG3}}
\end{figure}
\end{center}

Figure \ref{fig:FIG3} (a) show the spinful electronic band structure of Cs$_{2}$MoCl$_{6}$ with FM phase by DFT+$U$ method \cite{dft+u}, and the Hubbard $U$ term for Mo is 0.33 eV obtained by the linear response method \cite{dft+u2}. 
After considering the FM structure with magnetic moment on the (001) surface and SOC, $M_{z}$ and $C_{4z}$ will still be preserved, while $M_{x,y}$ and $C_{4x,4y}$ will be broken. Thus, the nodal chain on the $M_{x,y}$ plane will open a gap but still leave a single nodal ring protected by the $M_{z}$ symmetry. 
However, the nodal ring will move slightly away from $W$ point due to the symmetry breaking, and cross the high-symmetry line $W_{1}-W_{2}$ as shown in Fig. \ref{fig:FIG3} (b). 
Furthermore, SOC will make a change on the band structure along $\Gamma$-Z direction, and generate a pair of Weyl points protected by the $C_{4z}$ symmetry, as shown by the red and green hexagons in Fig. \ref{fig:FIG3} (b). 
Therefore, the $t_{2g}$ bands will be significantly affected after considering the FM order and SOC, which make the nodal chains into a pair of Weyl points and a single nodal ring, as shown in Fig. \ref{fig:FIG3} (b). 

\paragraph*{$k\cdot p$ model analysis}
Since two bands forming the Weyl points and a nodal ring are well separated from other bands, an effective $2\times 2$ Hamiltonian {based on the basis set \{$d_{xz}$, $d_{yz}$\}} can be constructed to describe the topological and other physical properties of the system. 
As discussed above, SOC and FM order will break the spin-rotation symmetry, $M_{x,y}$ and $C_{4x,4y}$ of the system, reducing the symmetry of the system to $C_{4h}$. 
Furthermore, the eigenvalues of the $\hat{C_{4z}}$ operator for those two bands along $\Gamma$-Z direction are $e^{-\frac{\pi i}{4}}$ and {$e^{\frac{-3\pi i}{4}}$}, so the $k\cdot p$ Hamiltonian describing the system around $Z$ with $C_{4h}$ double group symmetry can be written to the lowest order in $\mathbf{k}$ as:
\begin{eqnarray}
H(k_x,k_y,k_z)=\left(
\begin{array}{cc}
m-|\bm{k}|^2 & \tilde{D}k_zk_+ \\
\tilde{D}k_z{k_-} & -m+|\bm{k}|^2
\end{array}
\right), \label{t1}
\end{eqnarray}
where $k_\pm=k_x\pm ik_y$ and $\mathbf{k}$ is a wavevector measured from the $Z$ point. $\tilde{D}$ and $m$ are positive constants. 
We derived this Hamiltonian from the irreducible representations of the two bands around $Z$ after proper rescaling of the wavevector (see the Supplementary Material for details \cite{supple}). 
{Eigenvalues of ${H}$ are $E=\pm\sqrt{(m-|\bm{k}|^2)^2+\tilde{D}^2k_z^2(k_x^2+k_y^2)}$, which form a pair of Weyl points and a single nodal ring  located at ($k_x$ = 0, $k_y$ = 0, $k_z$ = $\pm\sqrt{m}$) and ($k_x^2+k_y^2=m$, $k_z=0$) with $E$ = 0, respectively, as shown in Fig. \ref{fig:FIG3} (c). }

As you may notice, the projections of topological band crossings on different surfaces will have different configurations. Namely, the two Weyl points will be projected inside of the nodal ring on the (001) surface, and to different sides of the projection of the nodal ring on the (010) surface (Fig. \ref{fig:FIG3} (c)). This difference will lead to different types of surface-state connection, which will be explored  
by both DFT calculation and model analysis in the following. Topological surface states obtained by DFT are calculated by the Green's function method \cite{sancho1984quick,wu2018wanniertools} with the tight-binding Hamiltonian obtained from maximally localized Wannier functions \cite{mostofi2008wannier90,wannier}. 

\paragraph*{Surface states of FM Cs$_{2}$MoCl$_{6}$ on the (001) surface}
Figure \ref{fig:FIG3} (d) shows the surface states on the (001) surface, where two Weyl points will be projected onto the center of the nodal ring, i.e., $\bar{\Gamma}$, as shown in Fig. \ref{fig:FIG3} (b). 
It is worth to mention that the topological surface states as well as the bulk bands are fully spin-polarized in Cs$_{2}$MoCl$_{6}$. 
In order to get a better understanding of the surface states connection and configuration, we also calculate the Fermi arcs with different energies, as shown in Figs. \ref{fig:FIG3} (g-h). 
The calculations show that the surface states on the (001) surface are in a drumhead shape, which is the characteristic of the nodal ring semimetal, and they will cross the projection of the Weyl points for systems with coexistence of Weyl points and a nodal ring. 
Thus, we can control the density of the surface states or the Fermi velocity of the topological surface states on the (001) surface by modulating the energy of Weyl points and nodal rings. 
In the following, we will also use the analytical model for a further discussion on the universal connection pattern between a pair of Weyl points and a nodal ring.

According to the $k\cdot p$ model in Eq.~(\ref{t1}), the nodal ring will be projected onto a circle $k_x^2+k_y^2=m$ and two Weyl points will be projected onto the $k_x=k_y=0$ point on the (001) surface.
In order to obtain the eigenvalues and eigenvectors of the surface states, we will deal with a semi-infinite system where the bulk and the vacuum lie in $z\leq0$ and in $z>0$, respectively.
In this case, $k_z$ is no longer a good quantum number, so we need to replace $k_z$ by $-i\partial_z$ \cite{bib1,bib2} and obtain the Hamiltonian of the semi-infinite system:
\begin{eqnarray}
H_z(k_x,k_y)=\left(
\begin{array}{cc}
m-R^2+\frac{\partial^2}{\partial z^2} & -i\tilde{D}Re^{i\theta}\frac{\partial}{\partial z} \\
-i\tilde{D}Re^{-i\theta}\frac{\partial}{\partial z} & -m+R^2-\frac{\partial^2}{\partial z^2}
\end{array}
\right) \label{t2}, 
\end{eqnarray}
 where $k_\pm=Re^{\pm i\theta}$ with $R(>0)$ and $\theta$ being real. 
By introducing a trial wavefunction proportional to $e^{\lambda z}$ and a parameter $\lambda$ satisfying Re($\lambda$)$>0$ to represent a surface state, we obtain the corresponding eigenvectors  
\begin{eqnarray}
\psi_{\pm}(z)=\left(
\begin{array}{c}
i\tilde{D}Re^{i\theta}\lambda_{\pm} \\
m-R^2+\lambda_\pm^2-E
\end{array}
\right)e^{\lambda_\pm z},
\end{eqnarray}
 where $\lambda_{\pm}^2=\frac{1}{2}\tilde{D}^2R^2-(m-R^2) \nonumber 
\,\qquad \pm\sqrt{E^2-(m-R^2)\tilde{D}^2R^2+\frac{1}{4}\tilde{D}^4R^4}$.
If we impose a boundary condition $\psi(z=0)=0$ for an eigenvector of the form $\psi=C_+\psi_++C_-\psi_-$ ($C_\pm$: constant), we obtain dispersionless DSSs with $E=0$, {and $\lambda_{\pm}$ becomes 
$\lambda_{\pm}=\frac{\tilde{D}R}{2}\pm\sqrt{\left(\frac{\tilde{D}R}{2}\right)^2+R^2-m}$} (see the Supplementary Material for details \cite{supple}). 
In order to meet the condition of Re($\lambda$)$>0$, $R$ should satisfy $|R|\leq\sqrt{m}$. 
Therefore, the DSSs cross the projection of the Weyl points lying at $E = 0$, which match with our DFT calculations very well. 

Next, we will study the penetration length for the surface states on the (001) surface, which have a unique connection between Weyl points and the nodal ring, by both the analytical model and DFT calculation separately. 
The penetration length $l$ for a wave function proportional to $e^{\lambda z}$ ($\lambda=\lambda_{\pm}$) is given by
 $l_\pm=\frac{1}{\mathrm{Re}(\lambda_\pm)}$. 
Here, we consider $l_-$ as the penetration depth because $l_-\geq l_+$.
When the surface state approaches the Weyl points or the nodal ring, the penetration length $l_-$ is expected to be divergent due to the bulk nature of these bands. 
Figure \ref{fig:FIG3} (e) shows the values of $\lambda_{\pm}$ as a function of $R$, which is between the projection of the Weyl points ($R=0$) and the nodal ring ($R=\sqrt{m}$). 
We notice that $R=\tilde{R}=\sqrt{\frac{4m}{\tilde{D}^2+4}}$ is a special point in that $\lambda_+=\lambda_-$ is satisfied and the behavior of $l_\pm$ changes drastically. 
{When $R>\tilde{R}$, $\lambda_\pm$ are real and the surface state decays into the bulk without oscillation. 
On the other hand, when $R<\tilde{R}$, $\lambda_\pm$ are complex and the surface states show an oscillating decay into the bulk. 
Further calculations obtained by a 47 layers' slab with the tight-binding Hamiltonian obtained from maximally localized Wannier functions} also show the same trend with the model calculation for both the real and imaginary part of $\lambda$, which are shown in Fig. \ref{fig:FIG3} (f).


\paragraph*{Surface states of FM Cs$_{2}$MoCl$_{6}$ on the (010) surface}

\begin{center}
\begin{figure}
\includegraphics[scale=0.8]{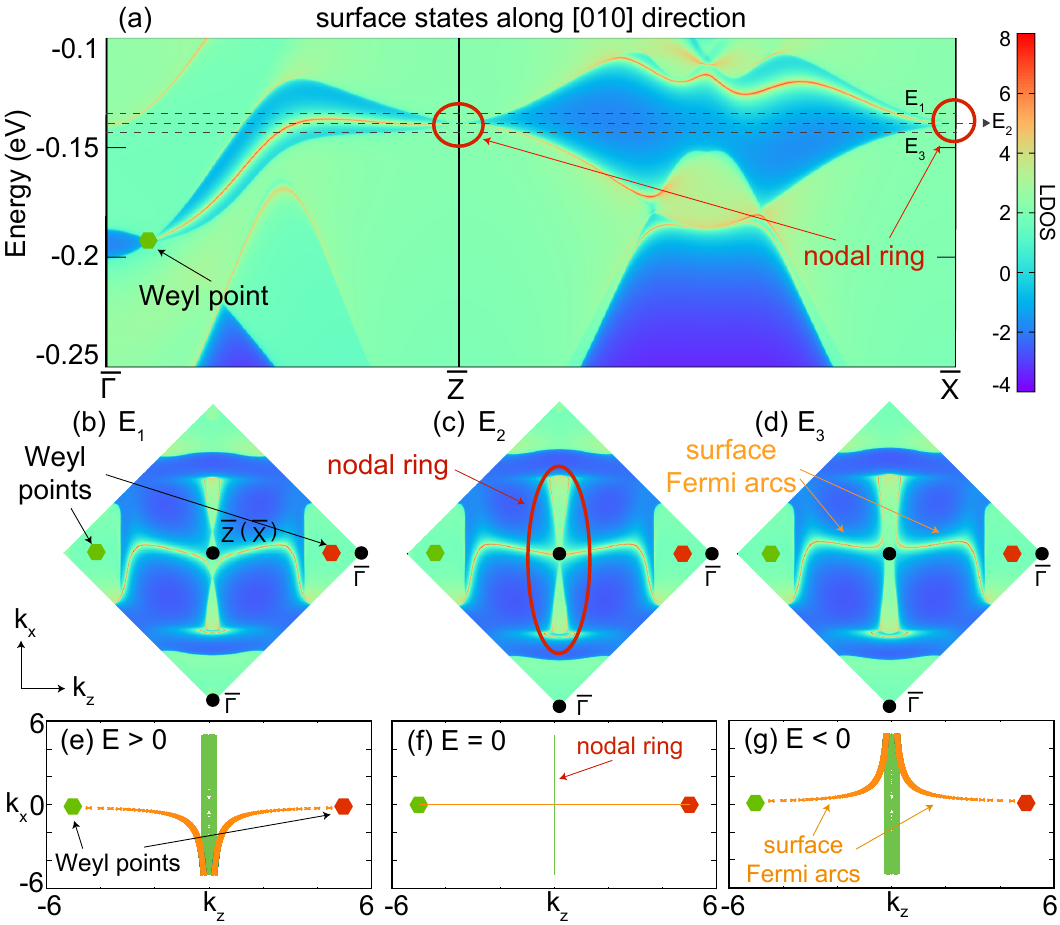}\caption{Surface states calculation on the (010) surface for Cs$_{2}$MoCl$_{6}$. (a) is the surface state on the (010) surface. (b-d) are the Fermi arcs with different energies labelled in (a), where $E_{2}$ corresponds to the energy of the nodal ring projected at $\bar{Z}$. 
(e-g) are Fermi arc calculations by the $k\cdot p$ model with $E$$>$0, $E$$=$0 and $E$$<$0. Parameters are $\tilde{D}=1$ and $\sqrt{m}=5$. 
The red and green hexagons represent the Weyl nodes with chiralities $+1$ (monopole) and $-1$ (antimonopole), respectively. 
\label{fig:FIG4}}
\end{figure}
\end{center}

On the (010) surface, two Weyl points will be projected on the $\bar{\Gamma}$-$\bar{Z}$ high-symmetry line, while the nodal ring will be projected to a line near $\bar{X}$ and $\bar{Z}$, as shown in Fig. \ref{fig:FIG3} (b-c). 
Figures \ref{fig:FIG4} (b-d) show the Fermi arcs with different energies labelled in Fig. \ref{fig:FIG4} (a), where $E_{2}$ corresponds to the energy of the nodal ring projected at $\bar{Z}$. 
When the energy for the isoenergetic contour decreases from $E_{1}$ to $E_{3}$, the Fermi arcs from Weyl points will cross through the projected points of the nodal ring, and rotate around two Weyl points. 
Thus, surface states on the (010) surface are in a helicoid shape and show the feature of Weyl points, which is different with the surface states on the (001) surface showing the feature of nodal ring. 

{This Fermi-arc surface states agrees with expectations from bulk quantities. 
At $\Gamma$ point ($k_{z} = 0$), the $\Gamma_{6}^{+}$ ($C_{4}=e^{-\frac{\pi i}{4}}$) irreducible representation has a lower energy than $\Gamma_{7}^{+}$ ($C_{4}=e^{-\frac{3\pi i}{4}}$), and at the $Z$ point ($k_{z} = \pi$), the $Z_{7}^{+}$ ($C_{4}=e^{-\frac{3\pi i}{4}}$) irreducible representation has a lower energy than $Z_{6}^{+}$ ($C_{4}=e^{-\frac{\pi i}{4}}$). 
From this band inversion between different $C_{4}$ eigenstates, we can derive the difference of Chern numbers ($Ch(k_{z})$) between $k_{z} = 0$ and $k_{z} = \pi$ subspaces. 
From Ref. \cite{fang2012multi,fang2012bulk}, we get $\frac{i^{Ch(k_{z}=\pi-\delta)}}{i^{Ch(k_{z}=\delta)}}=\frac{e^{-\frac{3\pi i}{4}}}{e^{-\frac{\pi i}{4}}}=-i$. 
Namely, $Ch(k_{z}=\pi-\delta)-Ch(k_{z}=\delta)\equiv -1$ ($mod$ 4). 
Here, $\delta$ is a small constant, which is introduced to make the spectrum gapped by avoiding the nodal ring. 
}
{This difference of Chern numbers is caused by the Weyl nodes at $\mathbf{k}=(0,0,\pm k_{0})$ ($k_{0}>0$), with a monopole charge $\mp 1$, respectively. 
In the present case, $Ch(k_{z})=0$ for $0<|k_{z}|<k_{0}$ and $Ch(k_{z})=-1$ for $k_{0}<|k_{z}|<\pi$, leading to topological chiral surface states with negative velocity $v_{x}$ on the (010) surface. It perfectly agrees with Figs. \ref{fig:FIG4} (b-d).
}

Further model calculations on the (010) surface also verify this point, as shown in Figs. \ref{fig:FIG4} (e-g) (see the Supplementary Material for details \cite{supple}). 
The projection of the Weyl points and the nodal ring are located at $(k_x,k_z)=(0,\pm\sqrt{m})$ and ($k_z=0$, $-\sqrt{m}\leq k_{x}\leq\sqrt{m}$), respectively.
When $E=0$ (Fig.~\ref{fig:FIG4}(f)), the Fermi arcs cross the projected nodal ring and connect two Weyl points straightly; when $E>0$ and $E<0$ (Fig.~\ref{fig:FIG4}(f) and (g)), Fermi arcs rotate around two Weyl points and cross through the nodal rings. 
Remarkably, when $E\neq0$, the Fermi arc meets the nodal ring tangentially, and the direction of the Fermi arc near the nodal ring change abruptly across $E=0$, as shown in Figs.~\ref{fig:FIG4}(e-g). 
It is due to an interplay between the helicoid nature of the surface state and the tangential property between the surface Fermi arc and a bulk Fermi surface. This abrupt change of the Fermi arc also appears in Figs.~\ref{fig:FIG4}(b-d) from the DFT. 
Both the DFT calculation and the model calculation on the Fermi arcs show perfect match with each other, which proves the correctness of our theory on the unique surface states connection. 

{We notice that such unique surface-state connection behavior can be found in other materials with different symmetries, such as in the electronic band structure of FM $HgCr_{2}Se_{4}$ (with SOC and without $\mathcal{T}$) \cite{xu2011chern}, the electronic band structure of FM double perovskites ($Ba_{2}NaOsO_{6}$, $Sr_{2}SrOsO_{6}$, $Ba_{2}ZnReO_{6}$, $Ba_{2}MgReO_{6}$, and $Ba_{2}CdReO_{6}$, with SOC and without $\mathcal{T}$) \cite{song2020symmetry,zhao2021coexistence}, and the phonon band structure of oxide perovskites ($BaTiO_{3}$, $PbTiO_{3}$, without SOC and with $\mathcal{T}$) }\cite{peng2020topological}. 
However, none of those works discuss on the connection of the unique surface states, and they mainly focus on the topology of the bulk bands. 
We demonstrate that both the drumhead surface states and the helicoid surface states will cross the projected points of the Weyl and nodal ring along different directions. Such surface-state connection for systems with the coexistence of Weyl points and nodal rings (and nodal lines) are universal, and it is also applicable to systems with/without considering SOC and $\mathcal{T}$.


\paragraph*{Conclusion}
We propose a FM topological half metal Cs$_{2}$MoCl$_{6}$ with spin-polarized bulk bands and surface states near the Fermi level. 
Due to the coexistence of Weyl points and a nodal ring, the surface states of Cs$_{2}$MoCl$_{6}$ show different properties along different directions. 
On the (001) surface, the surface states are in the drumhead shape, and it always go across the Weyl nodes. 
On the (010) surface, the Fermi-arc surface states are in a helicoid shape, and they meet the nodal ring tangentially, with their shapes change abruptly as a function of the energy. 
Analytical model calculation also match with the DFT calculation on both and surface states connection and configuration, which helps to verify our theory on the topological surface states. 
Such unique feature of the surface states connection is universal for systems with the coexistence of Weyl points and nodal rings, irrespective of the presence/absence of time-reversal symmetry and spin-orbit coupling. 


%
%
\paragraph*{Acknowledgements}
{We acknowledge the supports from JSPS KAKENHI Grants No. JP18H03678 and No. JP20H04633, Tokodai Institute for Element Strategy (TIES) funded by MEXT Elements Strategy Initiative to Form Core Research Center. T. Z. also acknowledge the support by Japan Society for the Promotion of Science (JSPS), KAKENHI Grant No. JP21K13865.}



\bibliographystyle{unsrt}
\bibliography{reference}
\newpage{}

\end{document}